\documentclass[10pt,aps,prl,reprint]{revtex4-1}

\usepackage[utf8]{inputenc}
\usepackage{graphicx}
\usepackage{amsmath}
\usepackage[colorlinks,pdfusetitle]{hyperref}
\hypersetup{colorlinks=true,allcolors=[rgb]{1,0.56,0}}

\newcommand{\orcid}[1]{\href{https://orcid.org/#1}{#1}}

\newcommand{\lsim}{\lesssim}
\newcommand{\gsim}{\gtrsim}

\newcommand{\eq}[1]{Eq.~(\ref{#1})}
\newcommand{\e}[1]{\times10^{#1}}

\newcommand{\ord}[1]{\mathcal{O}{(#1)}}
\newcommand{\beq}{\begin{equation}}
\newcommand{\eeq}{\end{equation}}
\newcommand{\msol}{M_\odot}
\newcommand{\rmP}{\bar M_{\rm P}}
\newcommand{\mP}{M_{\rm P}}

\begin{document}

\title{Connecting the Extremes:\\A Story of Supermassive Black Holes and Ultralight Dark Matter}
%\title{Supermassive Black Holes, Ultralight Dark Matter, and Gravitational Waves from a First Order Phase Transition}

\author{Hooman Davoudiasl}
\email{hooman@bnl.gov}
\thanks{\orcid{0000-0003-3484-911X}}

\author{Peter B.~Denton}
\email{pdenton@bnl.gov}
\thanks{\orcid{0000-0002-5209-872X}}

\author{Julia Gehrlein}
\email{jgehrlein@bnl.gov}
\thanks{\orcid{0000-0002-1235-0505}}

\affiliation{High Energy Theory Group, Physics Department, Brookhaven National Laboratory, Upton, NY 11973, USA}

\begin{abstract}
The formation of ultra rare supermassive black holes (SMBHs), with masses of $\ord{10^9\,\msol}$, in the first billion years of the Universe remains an open question in astrophysics. At the same time, ultralight dark matter (DM) with mass in the vicinity of $\ord{10^{-20}~\text{eV}}$ has been motivated by small scale DM distributions.  Though this type of DM is constrained by various astrophysical considerations, certain observations could be pointing to modest evidence for it.  We present a model with a confining first order phase transition at $\sim 10$ keV temperatures, facilitating production of $\ord{10^9\,\msol}$ primordial SMBHs.  Such a phase transition can also naturally lead to the implied mass for a motivated ultralight axion DM candidate, suggesting that SMBHs and ultralight DM may be two sides of the same cosmic coin.
We consider constraints and avenues to discovery from superradiance and a modification to $N_{\rm eff}$.
On general grounds, we also expect primordial gravitational waves -- from the assumed first order phase transition -- characterized by frequencies of $\ord{10^{-12}-10^{-9}~\text{Hz}}$.  This frequency regime is largely uncharted, but could be accessible to pulsar timing arrays if the primordial gravitational waves are at the higher end of this frequency range, as could be the case in our assumed confining phase transition. 
\end{abstract}

\maketitle

\section{Introduction}

The discovery of quasars --  believed to be powered by supermassive black holes (SMBHs) -- at redshift $4<z<5$
\cite{Turner1991} and at $ z > 7$ \cite{Haiman:2000ky,Wang_2021} --  prompts the question ``How did the first SMBHs grow so large so fast?''  It is possible that the formation of such SMBHs with mass $\gsim \mathcal{O}(10^9 M_\odot)$ -- the more distant cousins of the M87*  imaged by the Event Horizon Telescope in 2019 \cite{EventHorizonTelescope:2019dse} -- is the result of mergers and accretion of matter over a long period of time.  However, in general, very efficient processes and special conditions are required to be maintained over several orders of magnitude of mass growth for the formation of these SMBHs in the early Universe \cite{Inayoshi:2019fun}.  Whether or not these circumstances can be feasible is not a settled issue.  Hence, the appearance of such SMBHs at high redshifts poses an open question.  One possible explanation is based on the primordial formation of black holes \cite{Zeldovic67,Hawking:1971ei} coming from large density fluctuations.  Alternative mechanisms  have also been proposed recently \cite{Feng:2020kxv} in the context of Dark Matter (DM) which serve as seeds to SMBH formation \footnote{See e.g.~\cite{2021PhRvD.103f3012P}; for an overview of other proposed mechanisms to generate SMBHs see Ref.~\cite{Inayoshi:2019fun}.}.

In this letter we will consider the possibility that a first order phase transition (FOPT) in the early Universe, before the matter-radiation equality era, provided the catalyst for the formation of horizon size primordial SMBHs (pSMBHs).  
We will discuss some general aspects of pSMBH formation in the appendix.
In order to eschew the need for very efficient accretion and other special astronomical requirements, we will assume that the pSMBHs were formed near the puzzlingly large masses $\sim 10^9 \msol$.  In general terms, the onset of a FOPT leads to a suppression of the pressure response of a plasma \cite{Boyd:1996bx,Borsanyi:2012ve}, which could significantly enhance the likelihood that a horizon scale over-density would collapse and form a black hole.  Hence, it is well motivated to associate the formation of primordial black holes with a  FOPT in the early Universe; see the appendix for more details.  The maximum mass in the collapse is set by the thermal energy contained in a horizon volume. A pSMBH of mass $M\sim 10^9 \msol$ has a size $R\sim M/\mP^2\sim 10^{19}$~eV$^{-1}$, where $\mP \approx 10^{19}$~GeV is the Planck mass. This sets the Hubble scale $H\sim T^2/\mP$, with $T$ the temperature in the radiation era, corresponding to the pSMBH formation at $T\lsim \ord{10~{\rm keV}}$.  Such energy scales are interesting for another seemingly unrelated reason, as we will discuss next.  

Ultralight bosons of mass around $\ord{10^{-20}}$~eV provide a possible candidate for DM \cite{Hu:2000ke}.
This type of DM can also potentially address certain features of cosmic matter distribution that pose a challenge to the weakly interacting cold DM paradigm \cite{Bullock:2017xww}, and this scenario can be probed with astrophysical observations of the Lyman-$\alpha$ forest, dwarf spheroidal galaxies, and ultrafaint dwarf galaxies, among others, as we will  discuss later.  
A natural theoretical candidate for ultralight DM \cite{Preskill:1982cy,Abbott:1982af,Dine:1982ah} can arise via spontaneously broken $U(1)$ Peccei-Quinn symmetries \cite{Peccei:1977hh,Peccei:1977ur}, {\it i.e.}~the axion $a$ whose mass is protected by a shift symmetry and can hence be quite light \cite{Weinberg:1977ma,Wilczek:1977pj}; for a review see e.g.~\cite{GrillidiCortona:2015jxo}.  String theory can typically provide the requisite ingredients for such axions to arise, with decay constants $f_a$ not far from the reduced Planck mass $\rmP\approx 2 \times 10^{18}$~GeV \cite{Svrcek:2006yi}. The axion mass $m_a$ is given by 
\beq
m_a \sim \frac{\mu_a^2}{f_a} \sim 10^{-20}~\text{eV}\, \left(\frac{\mu_a}{\rm~keV}\right)^2\,\left(\frac{10^{17}~{\rm GeV}}{f_a}\right)\,.
\label{ma}
\eeq
The requisite small mass scale $\mu_a\sim$~keV  for an ultralight axion maybe generated by certain gravitational instantons, or else it may arise due to dynamics at low scales \cite{Hui:2016ltb}, in analogy with the QCD axion \cite{Davoudiasl:2017jke,Diez-Tejedor:2017ivd}. 
We will follow the latter path in this work, as implemented via a simple model described below \footnote{See also recent work in Ref.~\cite{Dvali:2021byy}, for a different approach to formation of SMBHs based on confinement.}.
In particular, we assume that the ultralight DM axion abundance is set by the misalignment mechanism with initial amplitude of oscillation of order $f_a$.
This mechanism can give the right relic abundance of DM for representative parameters such as those in \eq{ma}, see e.g.~Ref.~\cite{Hui:2016ltb}.

\section{A Specific Model Example}

%To examine the feasibility of the required ingredients for our scenario, here we will introduce a specific model.
Let us consider a dark $SU(3)_d$ gauge symmetry with one generation of heavy, vector-like dark quarks $\Psi$ \footnote{This choice of group corresponds to the minimal additional field content needed  to  fulfill the requirement of a FOPT, which is only present for $SU(N),~N\geq3$.
Larger groups would lead to larger $\Delta N_{\rm eff}$ effects, and likely a larger gravitational wave effect.}.
The quarks are charged under $SU(3)_d$ but are SM singlets and $\Psi_{L,R}$ have mass $m_\Psi\sim f_a\gg\mu_a$ such that we are in a regime similar to ``pure QCD" -- {\it i.e.}, in the quenched limit \footnote{In the quenched limit axions still gain a mass from instanton effects \cite{Rennecke:2020zgb}.} -- which can undergo a first-order confinement phase transition \cite{Borsanyi:2012ve}.  
Additionally, the PQ charges of $\Psi_{L,R}$ are such that they
are anomalous under the PQ symmetry which gives rise to the desired coupling of the axion to the dark gluons 
$\mathcal L\supset(a/f_a)G_{d\,\mu\nu}{\tilde G_d^{\mu\nu}}$, where $G_{d\,\mu\nu}$ is the dark gluon field strength tensor and $\tilde G_{d\,\mu\nu}$ is its dual.

The dark gluons will be present in the early Universe and they will contribute to the relativistic degrees of freedom during both Big Bang Nucleosynthesis (BBN) and the Cosmic Microwave Background (CMB) eras.
The constraint from the CMB on the effective number of neutrinos is slightly tighter at $N_{\rm eff}=2.99\pm0.17$ \cite{Planck:2018vyg} compared to the theoretical prediction of 3.045 \cite{deSalas:2016ztq} and yields a $2\sigma$ upper limit on $\Delta N_{\rm eff}<0.285$.
Hence, we will need to assume that the dark sector is at a somewhat lower temperature $T_d$ compared to the SM radiation temperature $T$.  The change in $N_{\rm eff}$ is given by 
\beq
\Delta N_{\rm eff} = \frac{4}{7}\left(\frac{11}{4}\right)^{4/3}
\left(\frac{T_d}{T}\right)^4 N_{\rm dG}\,,
\label{delNeff}
\eeq
where $N_{\rm dG}=N^2-1=8$ for the model adopted here.  
Using \eq{delNeff}, we then find $T_d\lsim 0.36\, T$, which suggests that the dark sector decoupled from the SM plasma well before BBN and subsequent transfers of entropy increased the SM sector temperature.

The dark gluons form glueballs upon confinement, at a temperature around $\mu_a$, where the lightest glueball has a mass $m_{\rm dGB}\gsim \mu_a$.  These bound states would dominate the energy density as DM unless there is a way for them to decay into dark or SM radiation.  In the SM, the QCD critical temperature $T_c\sim 160$~MeV \cite{Bhattacharya:2014ara} and the lightest glueball 
has a mass estimated to be $\sim 10 T_c$ \cite{Athenodorou:2020ani}.  Hence, we expect that the dark glueball population after confinement in the dark sector is characterized by a non-relativistic population of scalars (we assume that the higher excitations quickly decay or annihilate into the lightest glueball state).

Based on the preceding discussion, let us take the dark sector temperature to be given by $T_d\approx T/\sqrt{10}$ for ease of numerical analysis.  After confinement, the energy density of the gluon gas is inherited by the glueball population and we have 
\beq
\rho({\rm dGB}) \approx N_{\rm dG} \left(\frac{T_d}{T}\right)^4 \rho(\gamma)\,,
\label{rhodGB}
\eeq
where $\rho(\rm dGB)$ and $\rho(\gamma)$ are the energy densities of the glueballs and the SM photons, respectively.
We then get $\rho(\rm dGB) \approx 8\times 10^{-2} \rho(\gamma)$.
The contribution of the neutrinos to the SM energy density is given by $\rho(\nu)\approx 0.7 \rho(\gamma)$ and hence we get $\rho(\rm dGB) \approx \epsilon\, \rho(\rm SM)$, where $\epsilon\approx5\times 10^{-2}$ and $\rho(\rm SM)$ is the total SM radiation density.  The energy density in dark glueballs redshifts like $T^3$.  Hence, if the dark glueballs are stable until the Universe cools by about a factor $\sim 10$, they would surpass the SM radiation, which redshifts as $T^4$, and become the dominant form of energy.

In light of the above discussion, we demand that the glueballs decay quickly after formation.  This would roughly corrspond to a Hubble scale set by $T\sim$~few keV, and hence $H\sim T^2/M_P \sim  10^{-21}$~eV.  There are potentially several ways this can be achieved that involve adding extra ingredients to our model.  Here, only to illustrate the possibility of realizing prompt decays for glueballs, we offer a minimal approach. 

Let us denote a scalar parity even glueball state by $\Phi$ and let $\phi$ be a scalar, possibly another axion from the multitude of candidates that may arise in string theory \cite{Svrcek:2006yi}, for example.  If the shift symmetry of $\phi$ is broken softly, we may have the interaction $\mathcal L\supset\mu_\Phi \,\Phi \,\phi^2$
which can lead to a decay width for $\Phi$, given by 
\beq
\Gamma(\Phi\to \phi\phi) \sim \frac{\mu_\Phi^2}{16\pi \,m_\Phi}\,,
\label{GammaPhi}
\eeq
where $m_\Phi = m_{\rm dGB}$.  Requiring that $\Gamma(\Phi)\sim 10^{-21}$~eV, with $m_\Phi\sim 100$~keV, yields $\mu_\Phi\sim 10^{-7}$~eV.  

The above interaction can possibly descend from an operator of the type $G_{d\,\mu\nu}G_d^{\mu\nu}\phi^2/\Lambda^2$, where $\Lambda$ is large compared to $m_\Phi$.  This operator can then lead to $\mu_\Phi \sim \mu_a^3/\Lambda^2$.  For $\mu_a\sim 10$~keV, we find $\Lambda\sim 3$~GeV, which is well above energy scales relevant to our preceding discussion, and in particular those of the BBN.  The above dimension-6 operator can mediate interactions that bring $\phi$ into thermal contact with the dark gluons at temperatures of $\ord{\rm MeV}$ or higher relevant to the BBN.  However, this will be one more bosonic degree of freedom and will not significantly affect the required value of $T_d/T$ in our discussion.  

As long as the mass of $\phi$ satisfies $m_\phi\ll$~eV, the $\phi$ population will redshift as ``dark radiation" and will not result in an unwanted era of early matter domination.  The energy density in $\phi$ is inherited from the dark gluon population which can be small compared to the SM energy density during the CMB decoupling era $T\lsim 1$~eV, for the assumed $\Phi$ decays rates $\gsim 10^{-21}$~eV.  Hence, we hold that the remnant dark glueball population does not pose a severe problem.  Yet, some excess above the standard $N_{\rm eff}$, roughly at $\ord{0.1}$ level, may be expected in our model.
In addition to $\Delta N_{\rm eff}$, this model has a number of other general predictions that are also potentially testable, as outlined below.      

\section{Ultralight DM constraints}

As this model predicts an ultralight DM candidate, we start the discussion about the phenomenology of the model with the constraints on ultralight DM and then discuss some hints in the data.
First, Lyman-$\alpha$ forest measurements \cite{Rogers:2020ltq} disfavor ultralight bosons with masses $\lesssim2\e{-20}$ eV.
Future observations with DESI \cite{DESI:2016fyo} are expected to allow for improvement on this constraint.

Next, spin down of SMBHs due to superradiance \cite{Penrose:1969pc} disfavors ultralight bosons with masses $\gtrsim7\e{-20}$ eV \cite{Stott:2018opm}.
As more SMBHs are found and their spins are accurately measured, this constraint too can continue to broaden.
Constraints from the size of smallest DM structures in the Universe lead to a lower limit on the DM wavelength which translate to a lowest bosonic DM mass of $\gtrsim 10^{-22}$ eV.

\begin{figure}
\centering
\includegraphics[width=0.45\textwidth]{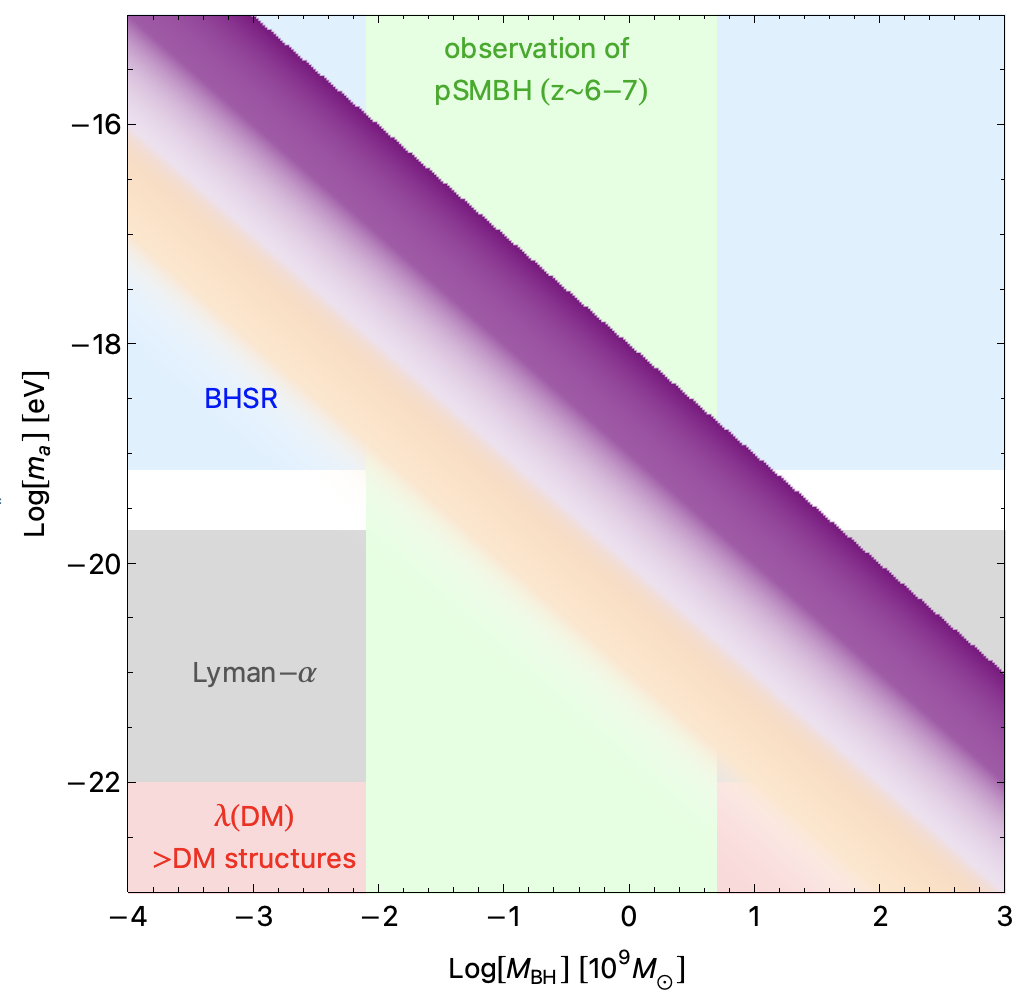}
\caption{Summary plot of the constraints and preferred regions of our model in the SMBH mass-axion mass plane. The green region shows the observed SMBH masses at redshift $\sim 6-7$. The blue region corresponds to constraints from BH superradiance (BHSR), the gray region shows the constraints from Lyman-$\alpha$ forest measurements, and in the red region the wavelength of  DM exceeds the smallest observed DM structures which provides a lower bound on the DM mass (see main text for more details). The orange and purple regions provide two benchmark scenarios for the relation between the axion mass and the primordial SMBH mass given in \eq{mamSMBH} with $f_a=10^{17}$ GeV (purple) and $f_a=10^{18}$ GeV (orange). The color intensity represents a decrease in $\varepsilon'$ from 1 to 0.01.  }
\label{fig:bounds}
\end{figure}

The existing bounds on both ultralight axion DM and pSMBHs are presented in Fig.~\ref{fig:bounds}.  Additionally, we show the relation between the axion mass and the BH mass. Here we have assumed that the relation between the size of the BH and the horizon scale is $R\sim \epsilon/H$
and that the temperature at BH formation  is given approximately by the energy scale of the FOPT $\mu_a\sim T$ such that the relation between the axion mass and the SMBH mass is
\beq
m_a=\varepsilon'\frac{\mP^3}{f_a M_{BH}}\,,
    \label{mamSMBH}
\eeq
where $\varepsilon'$ encompasses deviations from the  correspondence  $\mu_a\sim T$, and $R\sim 1/H$. In Fig.~\ref{fig:bounds} we show two benchmark points for $f_a \in \{10^{17},~10^{18}\}$ GeV  and $\varepsilon'$ between 0.01 and 1. For both benchmark points we find allowed regions which can explain the observed pSMBH population with axion masses not constrained yet.
The compatible region for $f_a$ may also be suggested by string theory \cite{Hui:2016ltb} as mentioned earlier.

There are several additional probes in this region of parameter space, each with its own theoretical uncertainties.
The first is from the size and age of the Eridanus II star cluster which disfavors DM with masses $\gtrsim10^{-19}$ eV \cite{Marsh:2018zyw}.
This constraint has recently been questioned with more involved numerical simulations in \cite{Schive:2019rrw} which indicated that Eridanus II could survive longer than previously thought in the presence of a soliton core.
Very recently, data from the center of the Milky Way was used to disfavor DM masses in the range $[10^{-20},3\e{-19}]$ eV \cite{Toguz:2021omv} which also seems to cover the relevant parameter space, although a complete picture with baryonic feedback may change this constraint.

The second is from an analysis of dwarf spheroidal galaxies (dSphs) which, on the surface, disfavors the parameter space in question, but also does not account for baryonic feedback \cite{Gonzalez-Morales:2016yaf} which likely modifies the dynamics of dSphs \cite{Read:2018fxs,Hayashi:2020jze}.
Since baryonic feedback is expected to be negligible for ultrafaint dwarf (UFD) galaxies \cite{Lazar:2020pjs}, we focus on these more robust environments for probing ultralight DM.

This leaves a tantalizing region of open parameter space.
Right in the middle of that parameter space is a hint for a finite wavelength for DM from the UFDs.
Reference~\cite{Hayashi:2021xxu} examined 18 UFDs and found that they prefer DM masses in the $\sim10^{-21}-10^{-20}$ eV region with considerable uncertainties.
We note that this is not yet at the level of discovery and the best fit point of the weighted average, $1.4\e{-21}$~eV, is disfavored by Lyman-$\alpha$ measurements.
Nonetheless, as these data sets continue to considerably improve in quality and quantity \cite{2021arXiv210809312M}, this is a prime target to test our model.

Measurements of the spin of a BH can be used to probe the physics of ultralight bosons via superradiance \cite{Arvanitaki:2009fg}. While this mechanism can be used to constrain their mass range \cite{Baryakhtar:2017ngi,Davoudiasl:2019nlo}, it is a challenging means for the discovery of ultralight bosons.
Nonetheless, it may still be possible to use superradiance to potentially identify the existence of an ultralight boson due to the formation of a cloud of particles surrounding the SMBH.
One mechanism that applies for ultralight axions is via a careful measurement of the polarization of light from the accretion disk \cite{Chen:2019fsq} around a SMBH with the correct mass by an experiment such as the Event Horizon Telescope \cite{EventHorizonTelescope:2019dse}.
In addition, as ultralight bosons enter the cloud in different angular momentum states a gravitational wave (GW) signature is formed \cite{Arvanitaki:2016qwi,Siemonsen:2019ebd} and the parameter space in question here could potentially be probed with observations of SMBHs by LISA \cite{Amaro-Seoane:2012aqc}.

\section{Gravitational Waves}

In addition to the possible GW signature from superradiance, several separate GW signatures may arise in this model.
With our assumption of a FOPT -- leading to the aforementioned dark glueballs -- a generic prediction of our model is the production of associated primordial GWs. These waves can be generated by true vacuum bubble collisions, sound waves, or magnetohydrodynamic turbulence (see, for example, Ref.~\cite{Caprini:2015zlo}). In the appendix, we provide an estimate of the expected amplitude and the frequency of the GW signal, from bubble collisions, which should yield roughly the right order of magnitude \cite{Schwaller:2015tja}, largely following the arguments presented in Ref.~\cite{Witten:1984rs}. For some more recent work on GW probes of phase transitions see, for example, Refs.~\cite{Grojean:2006bp,Caprini:2007xq,Caprini:2009yp,Croon:2018erz,Caprini:2019egz}. Production of GWs in confining phase transitions, over a range of energy scales, has also been discussed, for example, in Refs.~\cite{Tsumura:2017knk,Bai:2018dxf,Helmboldt:2019pan,Croon:2019iuh,Liu:2021svg}.
We also note that some care may be required when calculating the GW signature from strongly coupled FOPTs \cite{Croon:2021vtc}.

For fast phase transitions, GWs generated by sound waves are enhanced by the ratio of the velocity of the transition $\beta$ over the Hubble rate in comparison to the other sources of GWs. Hence, we will focus on sound wave GWs in the following. A consequence of fast phase transitions is also that GWs can only be sourced over a period shorter than a Hubble time.
In this case the energy density of the GW is \cite{Caprini:2015zlo,Ellis:2018mja,Ellis:2019oqb}
\begin{multline}
    \Omega_{sw}^{fast} h^2 = 8.53 \times 10^{-6}\left(\frac{H_*}{\beta}\right)\left(\frac{\kappa \alpha}{1+\alpha}\right)^2 \left(\frac{3}{g_*}\right)^{1/3}\\
   \times v_b \,S_{sw}\tau_{sw}H_*\,,
\end{multline}
where the spectral shape is given by 
\begin{equation}
   S_{sw}=\frac{f^3}{f_{sw}^3}\left(\frac{7}{4+3\frac{f^2}{f_{sw}^2}}\right)^{7/2}\,.
\end{equation}
The peak frequency is obtained from 
\begin{equation}
    f_{sw}=1.1\times 10^{-12}~\text{Hz}~\frac{1}{v_b}\left(\frac{\beta}{H_*}\right)\left(\frac{T_*}{10\text{ keV}}\right)\left(\frac{g_*}{3}\right)^{1/6}\,,
\end{equation}
with temperature of the phase transition $T_*$ and the number of relativistic degrees of freedom $g_*\approx 3$ at $T_*\approx 10$~keV.  For non-runaway bubbles with large wall
velocity $v_b \lesssim c$ an estimate for the efficiency factor $\kappa$ is 
\begin{equation}
    \kappa\approx \frac{\alpha}{0.73+0.083\sqrt{\alpha}+\alpha}\,,
\end{equation}
which depends on the energy released during the phase transition $\alpha$.
Phenomenologically $\tau_{sw}H_*\sim \mathcal{O}(10^{-3})$ \cite{Helmboldt:2019pan} for models with strong phase transitions and $\beta/H_*$ can be much larger than one; $\beta/H_*\approx \mathcal{O}(10^{4})$ is possible  \cite{Helmboldt:2019pan}.
For our numerical analysis we additionally assume $\alpha<1$ as the dark sector only contains a fraction of the total energy density of the Universe.

\begin{figure}
\centering
\includegraphics[width=0.5\textwidth]{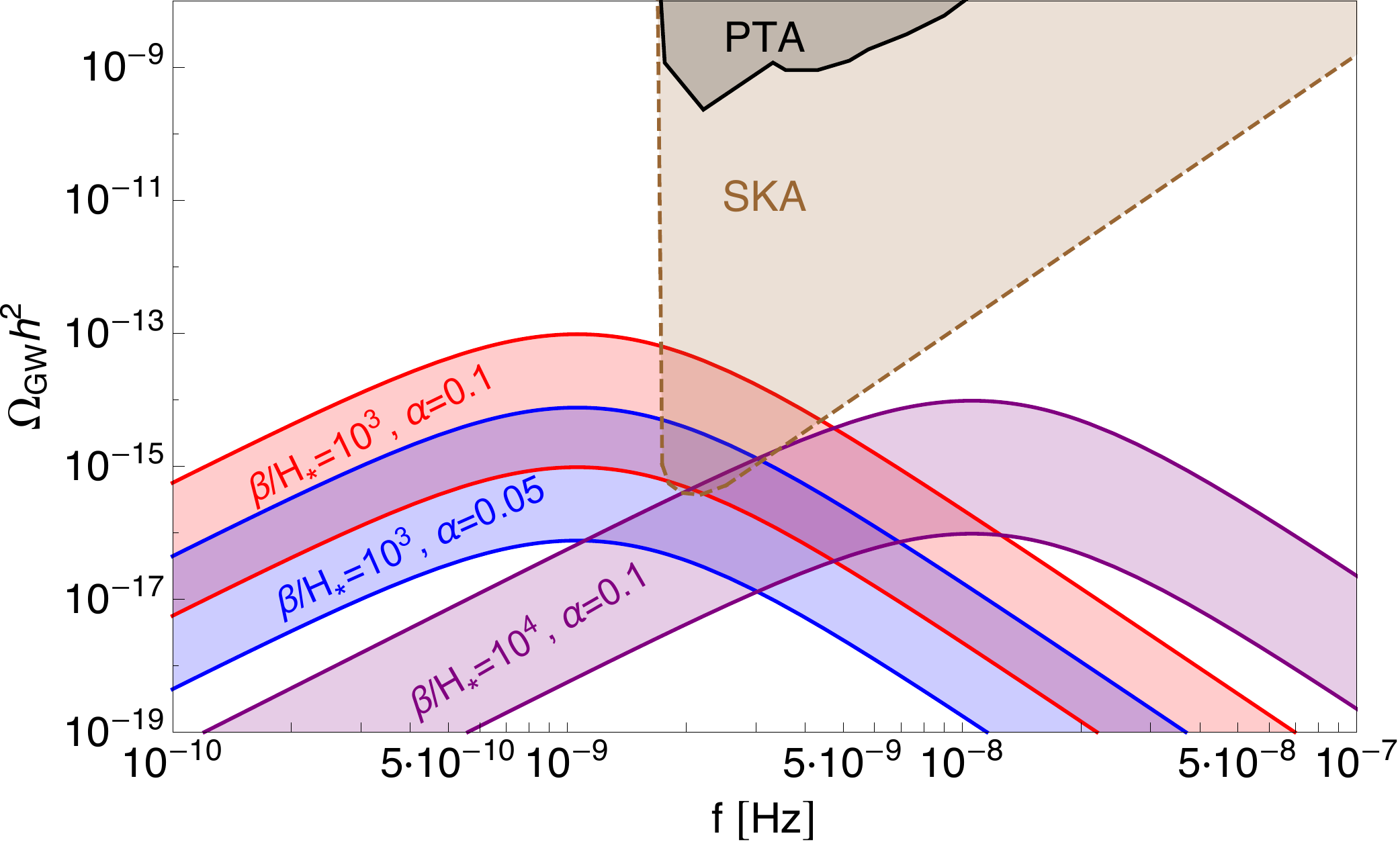}
\caption{Summary plot of the gravitational waves predicted in our model compared to existing experimental results and future sensitivities from pulsar timing arrays PTA (dark gray) \cite{Lentati:2015qwp,Shannon:2015ect,Aggarwal:2018mgp} and SKA (light brown) \cite{Zhao:2013bba,Moore:2014lga}, respectively. 
The predictions of our model are shown in  color for different values of $\beta/H_*$ and $\alpha$ with $g_*=3,~v_b=1, ~T_*=10$ keV.
The width of the bands corresponds to $\tau_{sw} H_* \in [10^{-3},~10^{-1}]$.
}
\label{fig:gw_bounds}
\end{figure}

In Fig.~\ref{fig:gw_bounds} we show the GW predictions from our model for benchmark values of parameters compared to current and future constraints from pulsar timing arrays.
We see that a part of our parameter space can be probed in the future with SKA \footnote{The SKA sensitivity may even be conservative as more suitable milli-second pulsars may be found \cite{Moore:2014lga}.} for $\beta/H_*~\sim 10^3$ and $\alpha\sim 0.05-0.1$.

We also note that NANOGrav has recently reported a hint of a GW signal that could be interpreted as a stochastic GW background \cite{NANOGrav:2020bcs}.  While the implied GW energy density is higher than our estimated model prediction, given 
the theoretically uncertain nature of the phase transition, the data could potentially be compatible with our scenario.

\section{Conclusions}

In this work, we presented a novel possible connection between the observations of $\sim 10^9 \msol$ supermassive black holes (SMBHs) at redshifts $z\sim 6-7$ and the possibility that dark matter (DM) is an ultralight axion of mass close to $\sim 10^{-20}$~eV.  The appearance of SMBHs at such an early epoch poses a puzzle.  Apart from being an interesting possibility, such axions may address certain small scale features of DM distribution and may be favored by measurements of ultrafaint dwarfs.  The connection that we propose is a dark sector confining first order phase transition, characterized by scales of $\ord{10~\text{keV}}$, that provides a catalyst for the primordial formation of $\sim 10^9 \msol$ SMBHs and endows the ultralight axion with mass, in a fashion similar to the QCD axion.

We confirmed that our model is consistent with a broad range of constraints.
Avoiding a large deviation from the standard number of relativistic degrees of freedom implies a cooler dark sector, but we typically expect a deviation at the $\ord{0.1}$ level as a consequence of our setup.  Also, generically, we expect gravitational waves in the pico to nano Hz regime, generated by the assumed phase transition.  This range of frequencies is not yet accessible to current measurements, but may be probed by pulsar timing arrays in the coming years. The presence of an ultralight boson at the suggested mass scales could result in SMBH superradiance signatures, adding potential extra support for our proposal.  A confirmation of this picture would establish ultralight DM and SMBHs as two sides of the same coin, and point to a new dynamical length scale in physics, corresponding to $\mathcal O(10$ keV$)$ energies, similar to the size of the hydrogen atom.

\begin{acknowledgments}
H.D.~thanks Brian Batell for  conversations on topics related to this work.  The authors acknowledge support by the United States Department of Energy under Grant Contract No.~DE-SC0012704.
\end{acknowledgments}

\bibliography{main}

\appendix

\section{Appendix}

\subsection{Formation of primordial supermassive black holes}
The formation of primordial black holes in the early Universe requires a density contrast $\delta$ to come into the horizon.  The mass of the black hole (BH) is then typically bounded by the horizon mass 
\beq
M_H \approx \frac{4}{3} \pi \rho H^{-3}\,,
\label{MH}
\eeq
at the time the perturbation crosses the horizon, where $\rho$ is the energy density corresponding to the Hubble scale $H$.  In general, one expects that there is a distribution of masses $\lsim M_H$ for the primordial BHs.  In our scenario, the dark sector makes up only a fraction of the radiation and hence the mass of the pSMBHs of interest are smaller than the horizon mass by a factor of $\lsim 0.1$, as will be discussed later. 

One can estimate the fraction $f_{\rm DM}$ of the cold DM (CDM) in the Universe composed of primordial BHs by (see, {\it e.g.},  Ref.~\cite{Byrnes:2018clq})  
\beq
f_{\rm DM} \approx \left(\frac{M_{eq}}{M_H}\right)^{1/2} \frac{\beta_i}{\Omega_{\rm CDM}}\,,
\label{f}
\eeq
where $M_H$ is the horizon mass at primordial BH formation and $M_{eq}\approx 3\times 10^{17}\msol$ \cite{Byrnes:2018clq} is the horizon mass at matter-radiation equality.  We will assume that the primordial BH mass is a fraction $\epsilon $ of $M_H$, and $\beta_i$ is the mass fraction of the Universe that ended up in primordial BHs at their formation during $i\in\{r,m\}$ for radiation or matter dominated epochs; $\Omega_{\rm CDM}\approx 0.27$ \cite{ParticleDataGroup:2020ssz} is the fraction of critical density in CDM. 

During a radiation dominated era, radiation pressure counterbalances the effect of $\delta$ that would otherwise facilitates collapse of the horizon energy content into a BH.  It is generally expected that the critical value of the density contrast required for collapse is $\delta_c\approx 0.45$ (see, for example, Refs.~\cite{Byrnes:2018clq,Carr:2020xqk}).
The literature indicates that $\delta_c$ takes values in the range 0.42 to 0.66 \cite{Polnarev:2006aa,Sato-Polito:2019hws}.
The value of $\beta_r$ in this era in given by \cite{Carr:1975qj}  
\beq
\beta_r \approx {\rm Erfc}\left(\frac{\delta_c'}{\sqrt{2}\,\sigma}\right)\,,
\label{beta-rad}
\eeq
where ``Erfc" denotes the complementary error function, and $\sigma$ is the dispersion in the density fluctuations.
The parameter $\delta_c'$, related to $\delta_c$ by $\delta_c'=\delta_c(1+\kappa\sigma/\delta_c)$, accounts for non-spherical collapses, where $\kappa=9/\sqrt{10\pi}$  \cite{Sheth:1999su,Sato-Polito:2019hws}. This correction makes the value of $\beta_r$ more suppressed compared to the case of spherical collapse.   

Since our first order phase transition will lead to a period of matter domination in the dark sector, $\beta_i$ can be much more enhanced \cite{Khlopov:1980mg}.
While lack of pressure facilitates the collapse into a BH, deviations from spherical symmetry can disrupt this process; this effect is parameterized by $\epsilon$. Following Ref.~\cite{Harada:2016mhb}, we roughly approximate the form of $\beta_m$ as
\beq
\beta_m \approx 0.1 \,(\epsilon \sigma)^5\,.
\label{beta-matter}
\eeq
As discussed in the context of a specific model in the main text, $\epsilon \sim 0.05$ can be a representative value for our purposes.

To explain the observations of SMBHs of mass $\sim 10^9\msol$, we need $\sim 100$ primordial BHs in this mass range \cite{Haiman:2000ky,Inayoshi:2019fun}.  Note that accretion is not expected to raise the masses of the pSMBHs significantly beyond their initial value at formation \cite{Carr:1974nx,Ricotti:2007au}.
The present total cosmic mass of DM is around $10^{22}\msol$, which suggests that the fraction in $\ord{100}$ pSMBHs of mass $\sim 10^9\msol$ would be $f_{\rm DM,9}\sim 10^{-11}$.  Using \eq{f}, we find $\beta_m \sim 7\times 10^{-16}$, which together with \eq{beta-matter} implies $\sigma \sim 0.03$ for fluctuations crossing the horizon at the epoch of pSMBH formation, corresponding to a matter dominated (zero pressure medium),  as proposed here. This same value of $\sigma$ corresponds to a  radiation era $\beta_r\sim 10^{-61}$, from \eq{beta-rad}, which is completely negligible.  Since $\beta_r\ll \beta_m$ for the regime of parameters typical of our study, the assumption of a FOPT is justified, as it enhances the pSMBH formation probability dramatically.        

Here, we would like to add a few comments.  In principle it is conceivable that these pSMBHs could form with the correct abundance in a radiation dominated environment without a FOPT, but this requires a larger value of $\sigma\sim 0.07$, following the above analysis.  The FOPT scenario can be even more favored, given a number of factors.

Firstly, the value inferred for $\sigma\sim 0.03$ in our FOPT (equivalent to matter domination) scenario can be lowered by around an order of magnitude if one can invoke models that do not require a very subdominant ($\epsilon\sim 0.05$) dark gauge sector component (this possibility could be realized if the dark gauge sector thermalizes after BBN and decays away after the FOPT; however, we do not give an explicit model here). 

Secondly, generating the requisite fluctuations in either scenario requires a jump in the density power spectrum $P_k$, from the strong CMB constraints \cite{Planck:2018vyg}, near our scales $k\sim (10~\text{kpc})^{-1}$ where $P_k$ is less constrained.
The jump required is large in both cases but is even larger for the case without a FOPT.  We note 
that constraints from pBH searches disfavor a power spectrum that corresponds to $\sigma\sim 0.07$, up to distance scales below but not very far from those relevant for our scenario \cite{Sato-Polito:2019hws}.  It may then be realistically anticipated that similar upper bounds may continue to remain comparable going to larger scales, corresponding to those in our scenario, as suggested by astronomical constraints on pBHs \cite{Carr:2020xqk}.  Thus a FOPT may not be a strictly necessary ingredient for pBH formation near our mass scales, but it can  potentially amplify the production considerably, and is hence quite well-motivated.

Our proposal connects the presence of high redshift SMBHs to the mechanism for ultralight axion DM mass generation.  As such, the properties of both sectors are tied by the energy scale of the FOPT, and hence the horizon scale, that leads to primordial BH production in the early Universe.
Here, we would also like to mention recent Ref.~\cite{Freitas:2021cfi} that considers a connection between ultralight bosons and SMBHs, however they employ a different approach based on the collapse of the DM clumps.

\subsection{Gravitational Waves from Bubble Collisions}
An upper bound on the wavelength $\lambda_*$ of the gravitational waves, produced at $T=T_*$, is given by the size of the horizon $R_*\sim \mP/T_*^2$, and hence 
\beq
\lambda_* \lsim \frac{\mP}{T_*^2}\,.
\label{lam*}
\eeq
This wavelength becomes stretched as the Universe expands and today it is given by 
\beq
\lambda_0 = \left(\frac{T_*}{T_0}\right)\lambda_*\,,
\label{lam0}
\eeq
where $T_0\approx 2.7$~K is the present temperature of the Universe. If bodies of horizon scale mass $\sim M_H$ have hard collisions near relativistic speeds, the gravitational waves produced are expected to have an amplitude of order the gravitational potential $\sim M_H/ (R_H \mP^2)\sim 1$, where $R_H$ is the horizon size. Assuming that the energy density in the dark sector is $\ord{\epsilon}$ of the SM radiation, we then expect the amplitude at $T=T_*$ to be given by $h_*\lsim \epsilon$. Note that this estimate scales with the square of the bubble size and could possibly be smaller \cite{Witten:1984rs} if the bubbles have sub-horizon scale.  Today, the amplitude of the primordial waves is given by \cite{Witten:1984rs} 
\beq
h_0 = \left(\frac{T_0}{T_*}\right) h_*\,.
\label{h0}
\eeq

Let us now estimate the numerical values of the above quantities for typical values of parameters assumed in our model.  Since we have a $T_d\sim T/3$, for $\mu_a \sim 10$~keV we have $T_*\sim 3$~keV and \eq{lam0} then yields $\lambda_0 \lsim 10^{23}$~cm, corresponding to a frequency of $\nu_0 \gsim 3 \times 10^{-13}$~Hz.  Using \eq{h0}, with $\epsilon\sim  5\times 10^{-2}$ as before, the amplitude of the GWs arriving at the Earth today is roughly given by $h_0\lsim 5\times 10^{-9}$.

The preceding estimate assumes horizon size bubbles, which is the maximal value.  Recent work \cite{Helmboldt:2019pan} suggests that the initial frequency of the GWs produced in a confining phase transition could be several orders of magnitude larger than this estimate.  Note that the smaller wavelength implies a reduced initial amplitude  \cite{Witten:1984rs}.  For this reason we focus on the expansion
of bubbles leading to sound waves as the dominant source of GWs here.

\end{document}